# Design and analysis of 5045 S-band klystron DC electron gun

Abdul Rehman[1], Munawar Iqbal[1,a)], Z. Zhou[2]

[1]Centre for High Energy Physics, University of the Punjab, Lahore 45590, Pakistan

[2]Institute of High Energy Physics, Chinese Academy of Sciences, Beijing 100049, China

## Abstract

The design and performance analysis of DC electron gun for 5045 S-band klystron has been worked out using SLAC beam trajectory program (EGUN) and Computer Simulation Technology Particle Studio (CST-PS), Codes. Specifications of electron gun were focused on beam; current, perveance, size and emittance. Optimized beam; current, perveance, diameter and emittance were 414.00A, 2.00μP, 26.82 mm and 103.10 π mm mrad, respectively. Furthermore, the optimized characteristic parameters of the gun were also calculated and compared with the simulated and experimental values which were in agreement. Accuracy of simulation was verified by comparison of emitted beam current which has error of zero percent.

[a)]Author to whom correspondence should be addressed.
Electronic mail: muniqbal.chep@pu.edu.pk

**Introduction**

High power klystron plays a vital role in variety of applications that include high-energy accelerators, long-range radar, and communications[1]. Medicine is another major and socially important klystron application[2]. It is a microwave amplifier through a DC electron beam to enhance the energy of the charged particle beam. A DC electron gun is a starting point and critical both for; klystron and accelerator, hence, is a fundamental key component of the accelerator technology. In klystron, the electron gun provides an electron beam in a suitable shape and intensity to interact with RF wave. Klystron's power gain majorly depends on the current and voltage of the electron beam[3]. The electron gun should supply desired beam current and beam quality to get linear, reliable and long lived klystron. So, the design and manufacturing of electron gun is important in making a klystron. Thermionic cathode is the most common electron gun[4] in klystron since thermionic mode of emission of electrons is easy and inexpensive. Modeling and simulation of beam dynamics with space charge effects are significant in contemporary research of particle accelerators and particle sources[5]. We used a working klystron example, with its actual performance in comparison to the results from two different codes. This is useful information, since, this gun has high perveance, analytic calculations give very slight assistance for high-perveance gun design; hence we have to use numerical methods[6]. We simulated the SLAC 5045 S-band Klystron's DC electron gun, as it has become a very successful RF source for the linear accelerators[7] by removing RF part (shown in figure 1), form the assembly using EGUN[8] and CST-PS[9] codes and then compared the results with the experimental values. The design of the gun was focused on an optimization of the structure for the emitting beam current, perveance, size and minimizing the emittance.

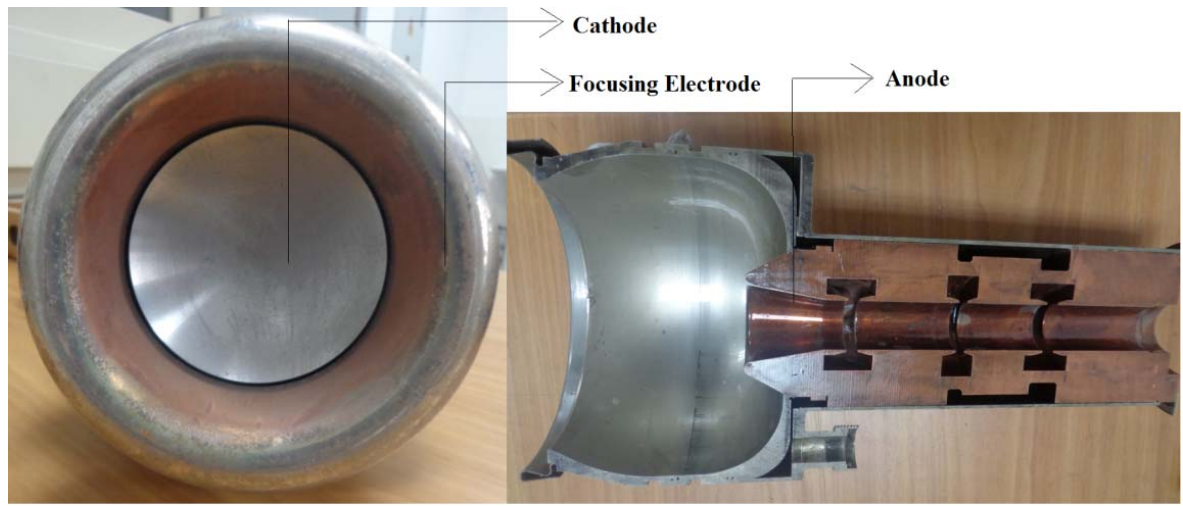

Fig. 1: Klystron's RF gun actual view

**Modeling & Simulation of the gun**

Geometrical patrameters of the gun were measured from the physical diemnessions and verified from SLAC engineering document. Table 1, describes the numerical values of mutual distances and potentials of electrodes.

Table 1: Geometrical & potential parameters of the gun

| Sr # | Electrodes' specification | Values |
|---|---|---|
| 1 | Cathode spherical radius (mm) | 73.69 |
| 2 | Cathode Radius (mm) | 53 |
| 3 | Cathode to anode distance (mm) | 44.51 |
| 4 | Focusing electrode to anode distance (mm) | 19 |
| 5 | Focusing electrode angle (degree) | $10^o$ |
| 6 | Anode aperture radius (mm) | 35 |
| 7 | Anode potential (kV) | 350 |
| 8 | Cathode potential (V) | 00 |
| 9 | Focusing electrode (V) | 00 |

Accodrding to the EGUN protocol, POL file was prepared from these geometrical parameters. Also, the gun was modelled in CST PS keeping the same geometrical input values as were used in EGUN. A cutaway and cathode surface modeled in the CST & EGUN is given in Figure 2.

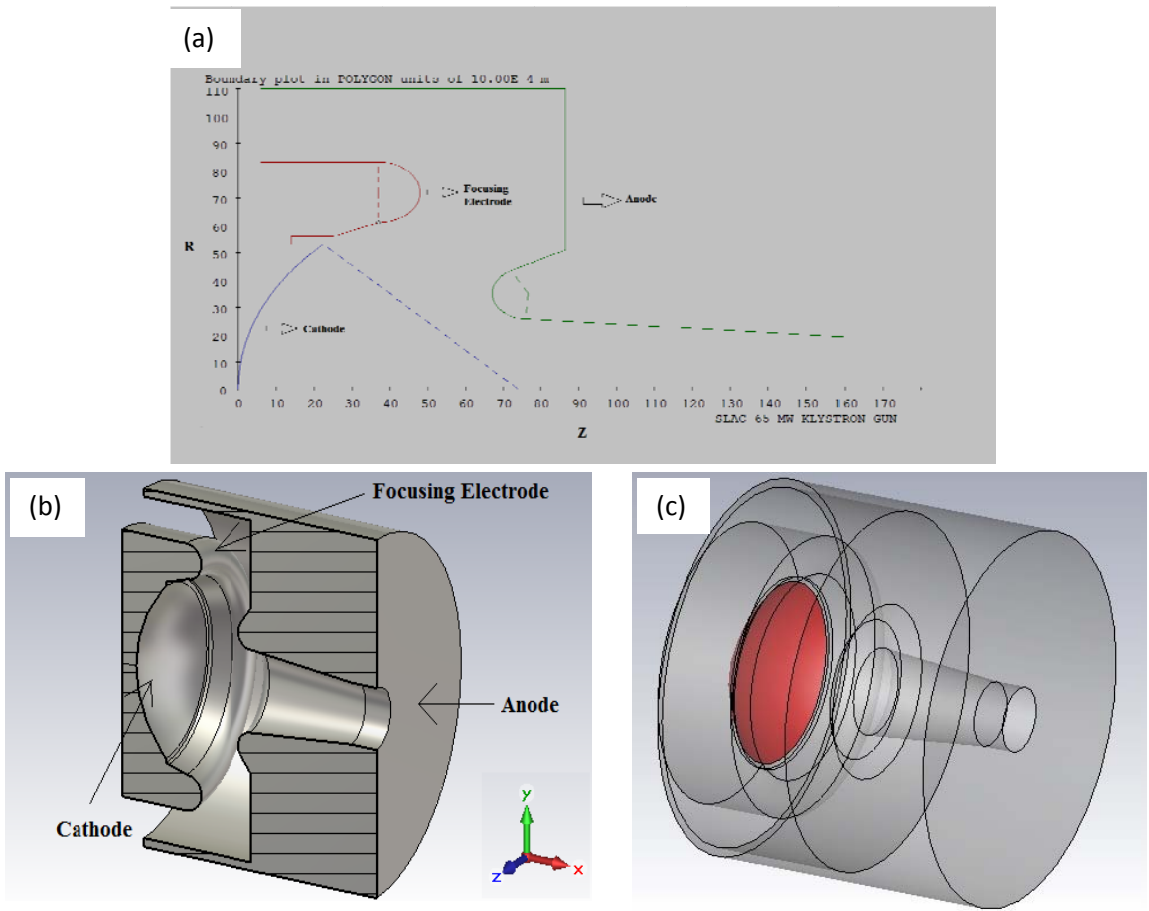

Fig. 2: Gun modelling: a) EGUN POLYGON b) CST PS c) emissioin surfacce highlighted in red (cathode)

## Results and discussions

Then, the gun was simulated using the two codes to calculate the beam trajetories that are shown in Figure 3.

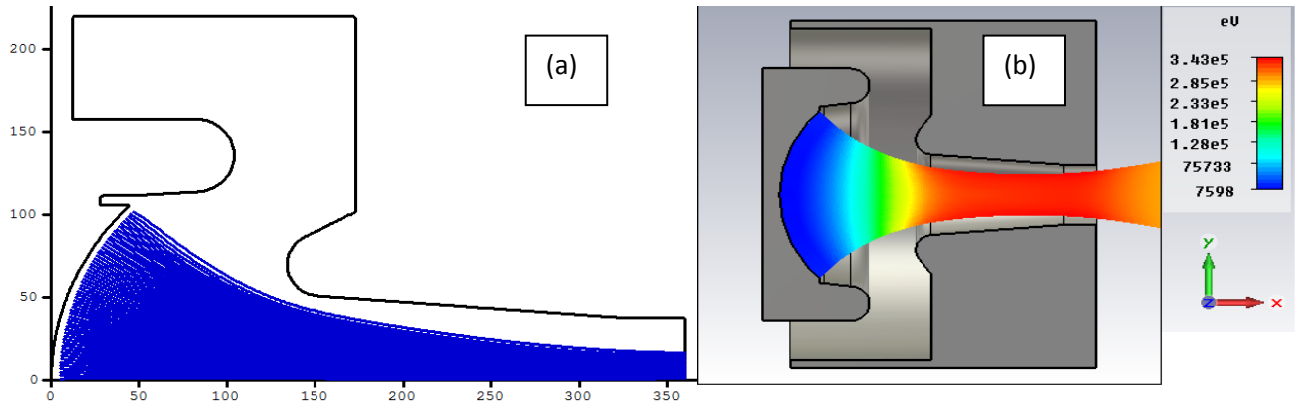

Fig. 3: Beam trajectories of the gun by a) EN2w b) CST PS

Beam trajectories were not colliding with electrodes' surface, therefore, no thermal load on the anode was observed. The beam current and perveance were measured to be 414A and 2 µP, respectively. The beam parameters were in agreement with the experimental values[7]. The potential, charge density, electric field and current density of the particles are shown in Figures 4 & 5, respectively.

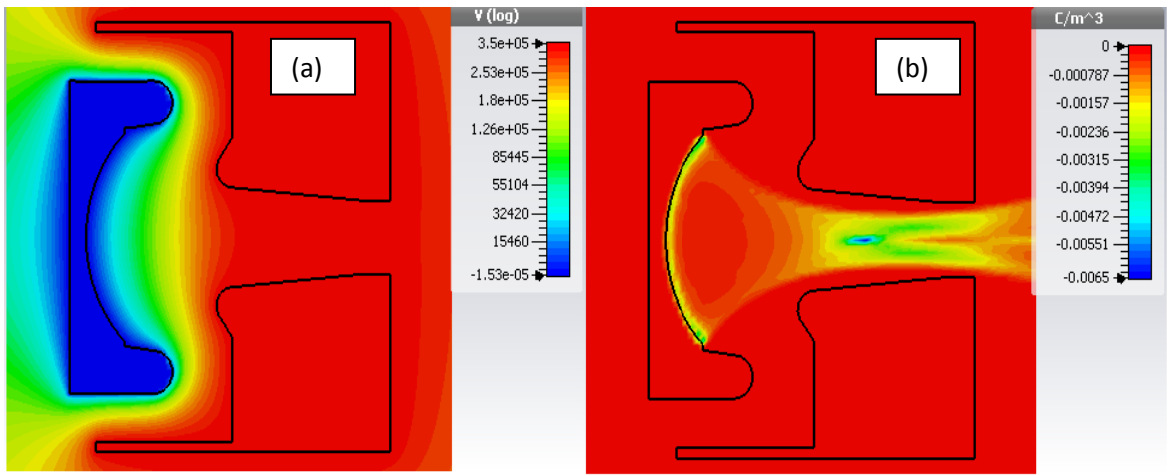

Fig. 4: a) Potential contours b) charge density

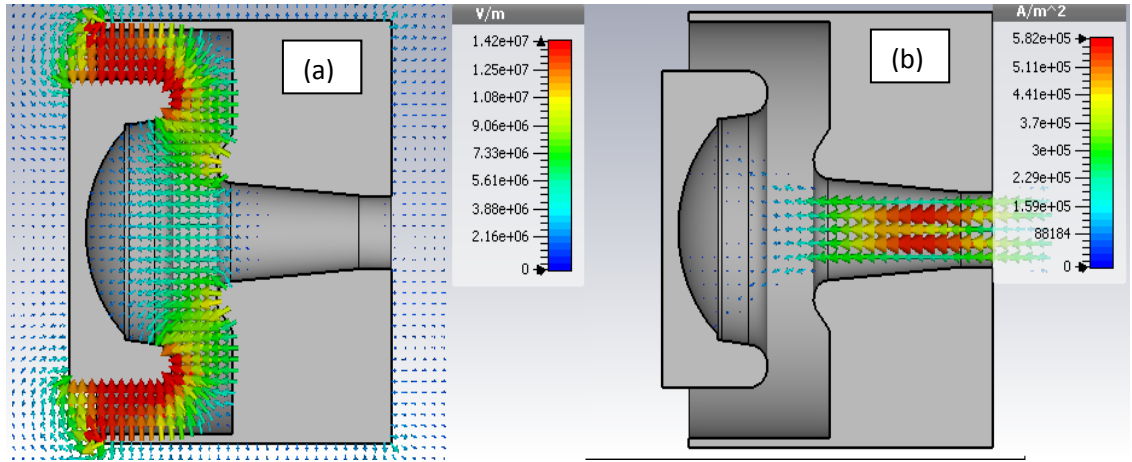

Fig. 5: a) Electric field lines b) Current density of the gun

Figure 6, represents the beam profile as it moves from cathode to anode region. Initially, the beam diameter was 90.770 mm at 30mm away from cathode's center in normal direction as is shown in Figure 6a. As, the gun anode aperture was smaller than the cathode diameter, therefore, focusing electrode was used to reduce the diameter of the beam. Beam's diameter was constant inside the anode as is seen in Figure 6 ("c" and "d"). After the beam waist (26.82mm at 157mm away from cathode, smallest beam diameter) beam again started diverging due to internal repulsion force. To make the beam diameter as per requirement of the other parts of klystron, solenoids (magnetic focusing) were used to focus the beam.

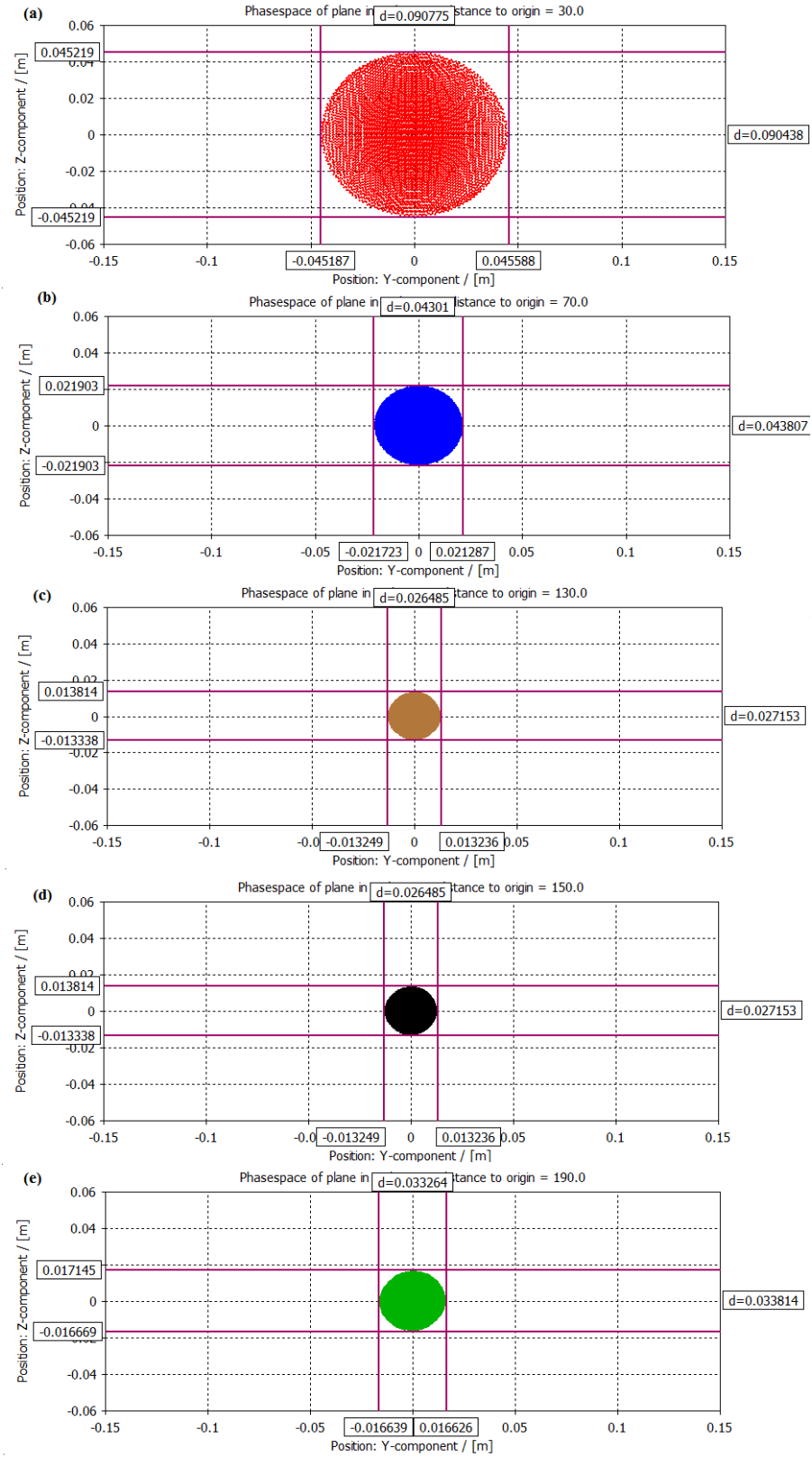

Fig. 6: yz-position along the propagation of the beam.

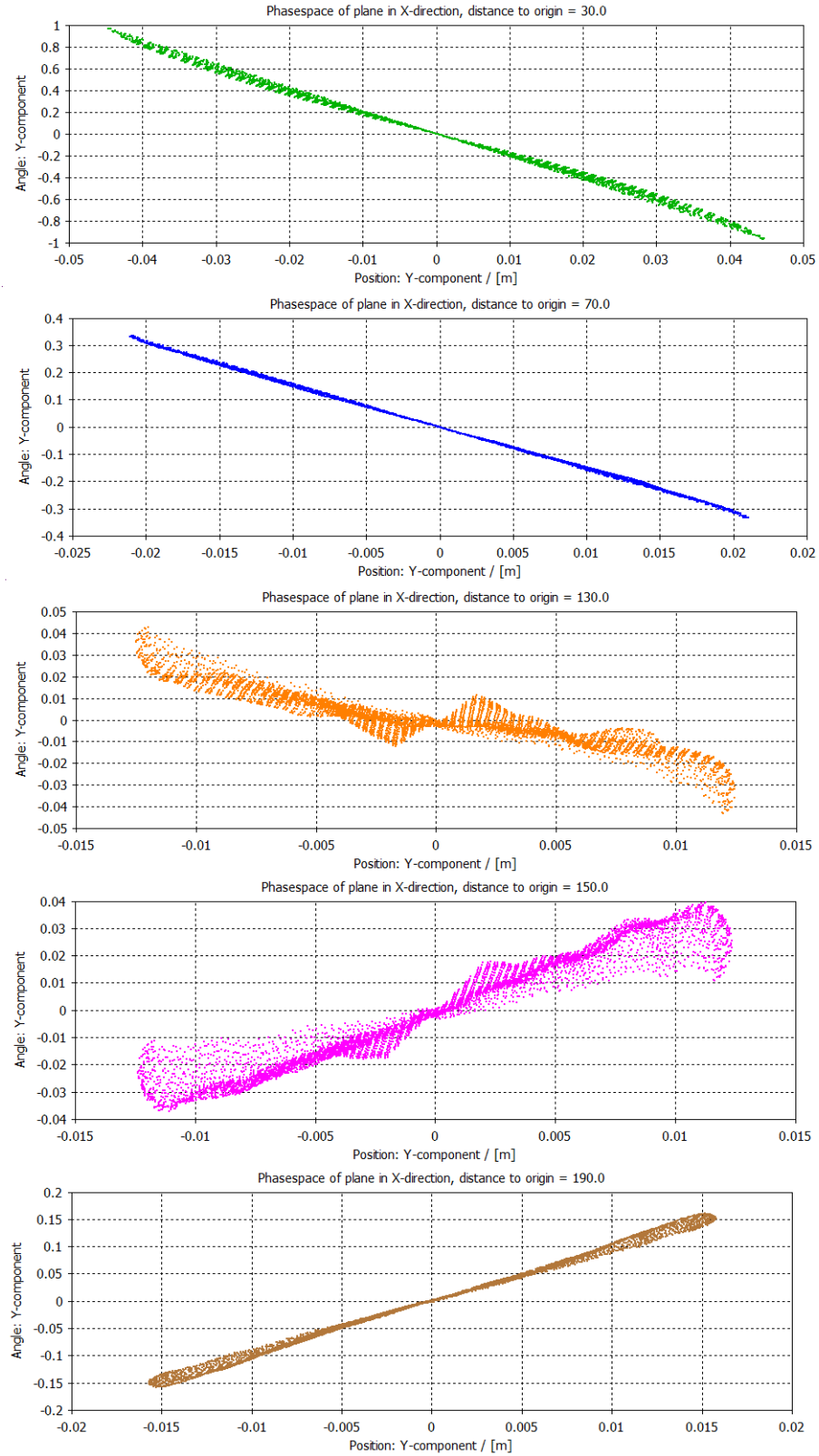

Fig. 7: Y-Phase space plot of plane in x-direction

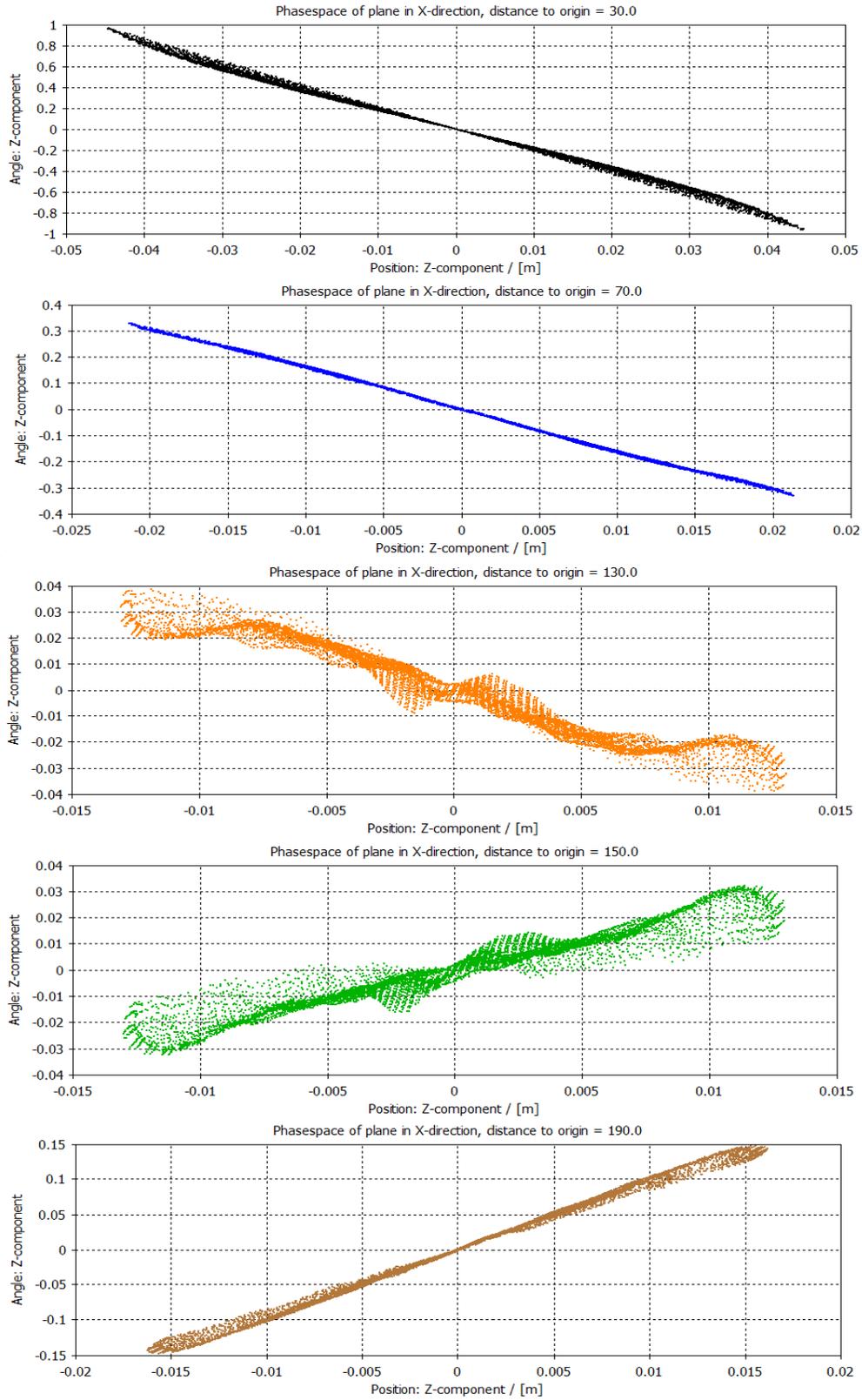

Fig. 8: Z-Phase space plot of plane in x-direction

The phase space plot displays the piercing angle of the particles' velocity to the monitoring plane's normal with respect to the position of the piercing point in one dimension. Choosing the same position direction and angle direction leads to a phase space-diagram, this clearly identifies the focusing or defocusing property of the beam at the monitor's position. Figures 7 & 8 are showing the y-phase space plots and z-phase space plots respectively.

Figures 9 and 10 show the transverse emittance (y and z emittance) variation at these particle monitors along the propagation of beam. These plots confirm the variation in diameter of the beam and reflect that beam is fully collimated in the anode region with emittance around 103.10 πmm-mrad.

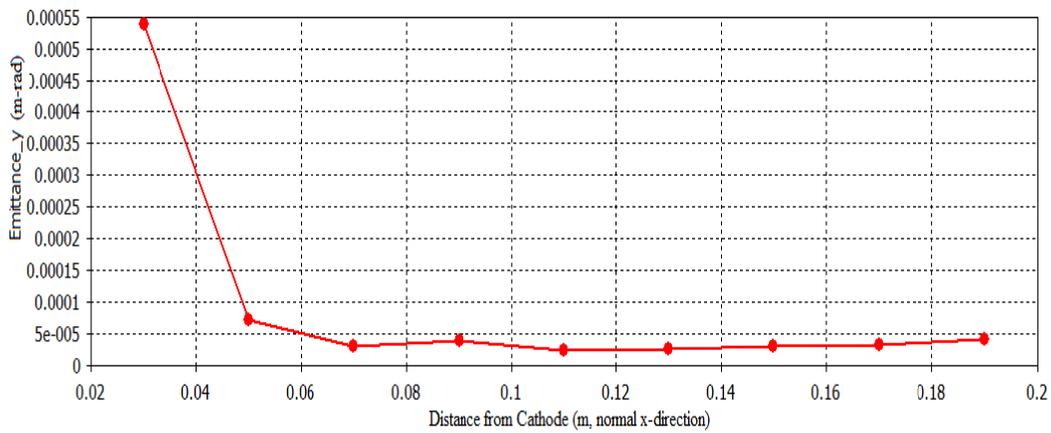

Fig. 9: Transverse y-emittance variation along the propagation of beam

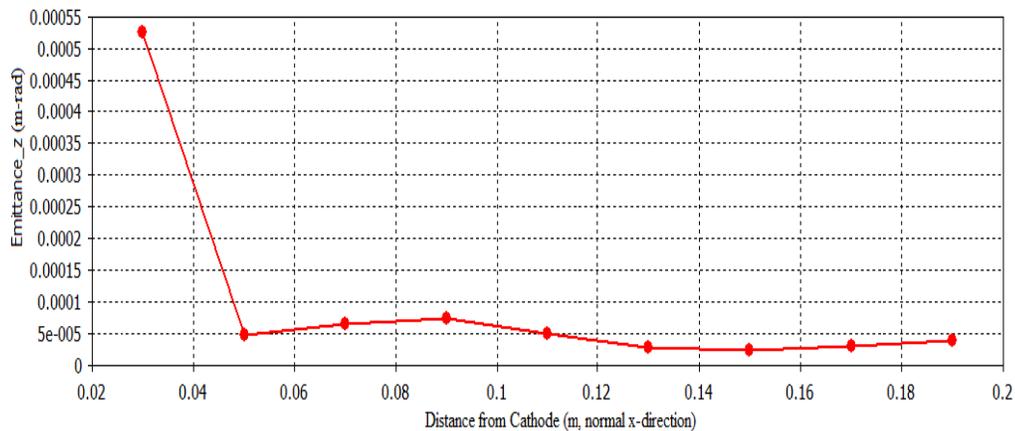

Fig. 10: Transverse z-emittance variation along the propagation of beam

This is a high perveance gun; therefore, perturbation is strong at anode and cathode electric fields. The Pierce gun theory[10] is not valid for high perveance guns. Modification on the focusing electrode can compensate large anode aperture effects. Since theory is not valid, simulation tool should be used to optimize high perveance gun design[11]. We changed three geometrical and one electrical parameter to obtain optimized beam parameters as close to experimental values and compared the beam current and perveance obtained from CST-PS and EGUN.

## Gun Parameterization

**Focusing electrode angle**

Focusing electrode angle has strong effect on beam dynamics; variation in focusing electrode angle changes the strength of electric field. Due to change in electric field, the beam current and its other parameters are changed. Table 2, shows the variation of focusing angle at cathode to anode distance=44.51mm, and anode aperture radius=35mm. Figure 11, shows the comparison of current (A) values at different focusing electrode angle, of EGUN and CST-PS, and confirms the experimental and simulated result. Figure 12, shows the beam trajectories at all focusing electrode angles.

Table 2: Focusing electrode angle effects

| Focusing electrode angle (deg) | Current(A) EGUN | Current(A) CST-PS | Perveance (μP) EGUN | Perveance(μP) CST-PS | Electric field Maximum (V/m)) CST PS |
|---|---|---|---|---|---|
| 70 | 802.291 | 802.29 | 3.87 | 3.87 | $1.15 \times 10^8$ |
| 60 | 726.28 | 726.29 | 3.50 | 3.51 | $4.8 \times 10^7$ |
| 50 | 678.5 | 678.4 | 3.27 | 3.26 | $4.2 \times 10^7$ |
| 40 | 599.3 | 599.28 | 2.89 | 2.88 | $4 \times 10^7$ |
| 30 | 524.5 | 524.55 | 2.53 | 2.54 | 3.99x10^7 |
| 20 | 448.02 | 448.09 | 2.16 | 2.17 | 3.85x10^7 |
| 10 | 414.02 | 414.01 | 2.00 | 2.00 | 3.1x10^7 |
| 0 | 352.5 | 352.54 | 1.70 | 1.71 | 3.39x10^7 |

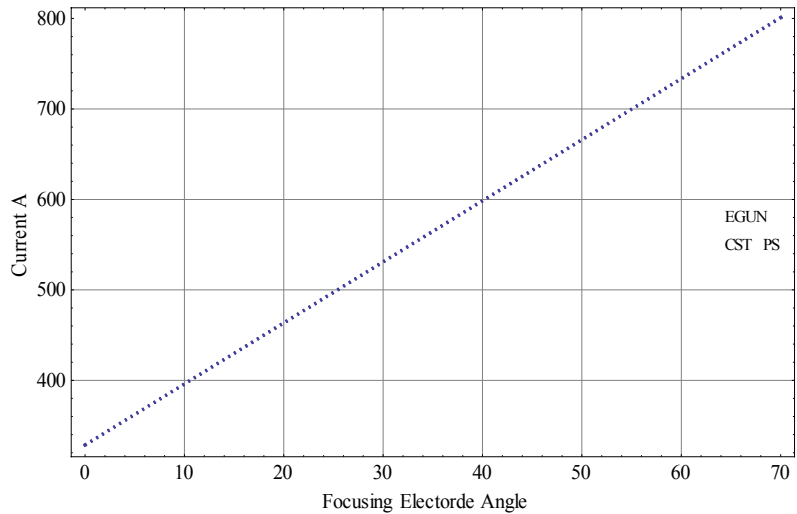

Fig. 11: Comparison of current (A) at different focusing electrode angles, of CST-PS and EGUN.

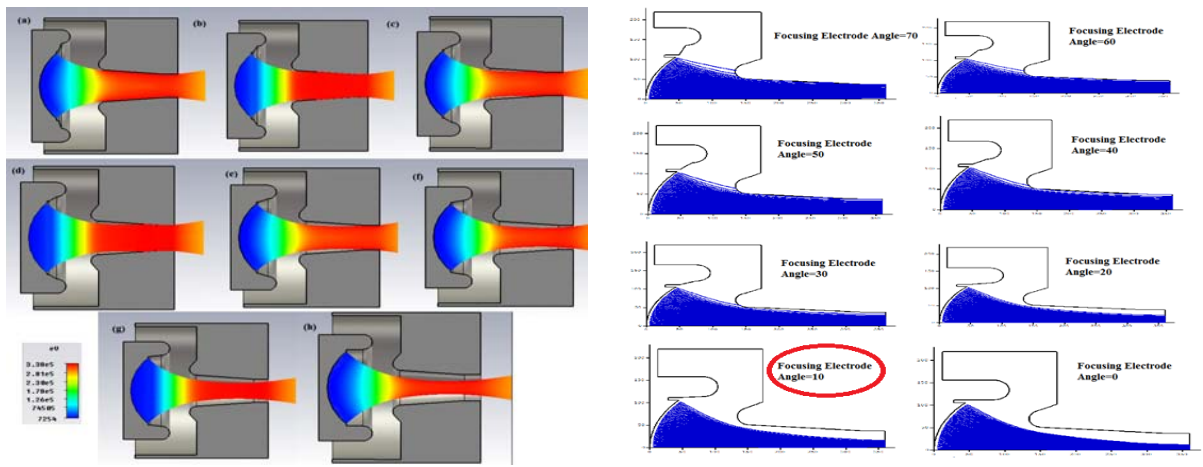

Fig. 12: beam trajectories at different focusing electrode angle by CST & EGUN: (a) 70°, (b) 60°, (c) 50°, (d) 40°, (e) 30°, (f) 20°, (g) 10°, (h) 0°

## Cathode to anode distance

Cathode to anode spacing has also very strong effect on beam dynamics. By varying the cathode to anode distance in EGUN & CST-PS, keeping constant focusing angle=10 and anode aperture radius=35mm, the effects on beam parameters are summarized in Table 3.

Table 3: Cathode to anode distance effects

| Anode to cathode Spacing (mm) | Current(A) EGUN | Current(A) CST-PS | Perveance ($\mu$ P) EGUN | Perveance ($\mu$P) CST-PS | Electric field Maximum (V/m)) CST PS |
|---|---|---|---|---|---|
| 14.51 | 3315.6 | 3315.62 | 16.01 | 16.012 | $5.53 \times 10^8$ |
| 24.91 | 1435.5 | 1435.489 | 6.93 | 6.921 | $3.99 \times 10^7$ |
| 34.51 | 745.32 | 745.34 | 3.59 | 3.6 | $3.17 \times 10^7$ |
| 44.5 | 414.0 | 414.0 | 2.00 | 2.00 | $3.16 \times 10^7$ |
| 54.51 | 306.4 | 306.6 | 1.47 | 1.48 | $3.12 \times 10^7$ |
| 64.51 | 231.62 | 231.63 | 1.11 | 1.112 | $3.05 \times 10^7$ |
| 74.51 | 189.9 | 189.89 | 0.91 | 0.90 | $3 \times 10^7$ |
| 84.5 | 165.7 | 165.71 | 0.8 | 0.801 | $2.92 \times 10^7$ |

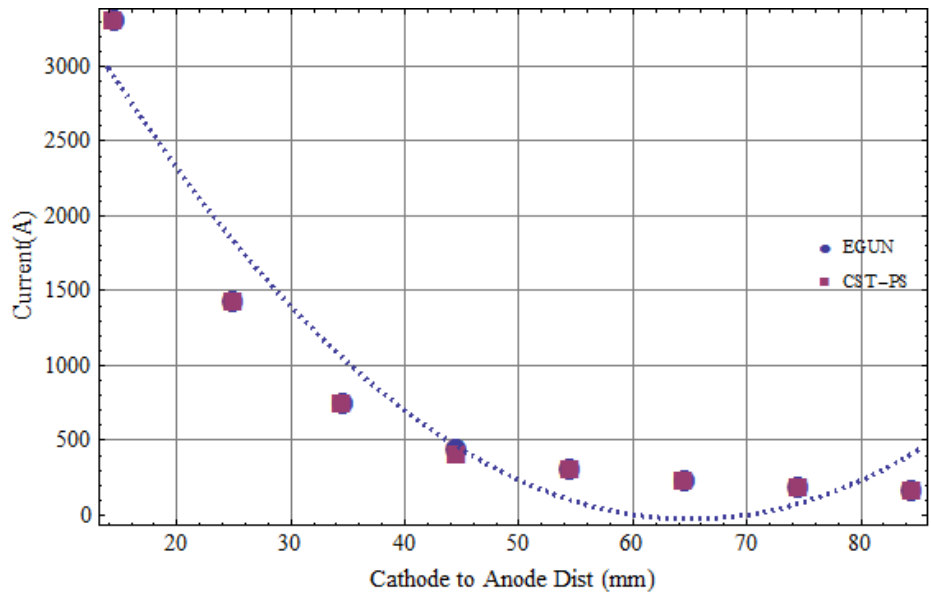

Fig.13: Comparison of current (A) at different cathode to anode distances, of CST-PS and EGUN

Figure 13 shows the comparison of current (A) values at different cathode to anode spacing, of EGUN and CST-PS, and confirms the results. Figure 14, shows the beam trajectories at different cathode to anode spacing.

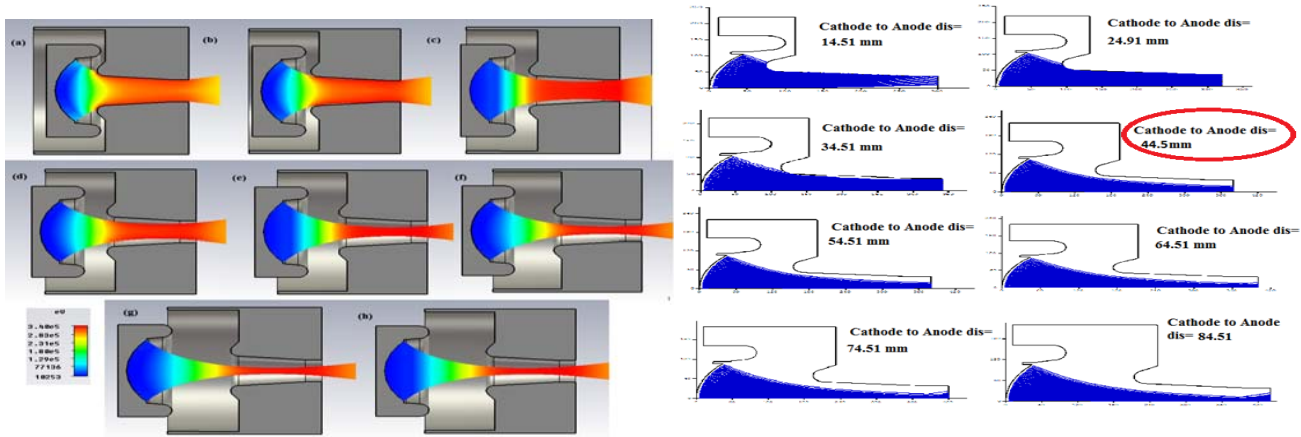

Fig. 14: Beam trajectories at different cathode to anode distances: (a) 14.51 mm , (b) 24.51 mm, (c) 34.51 mm, (d) 44.51 mm, (e) 54.51 mm, (f) 64.51 mm, (g) 74.51 mm, (h) 84.51 mm

**Anode aperture radius**

Anode aperture radius has not a dreadful effect on beam dynamics, but the current and other parameters change with the variation in anode aperture radius. By keeping the fixed focusing angle=10 and cathode to anode distance = 44.51mm, the anode aperture radius and its effects on beam parameters are listed in Table 4.

Table 4: Effects of anode aperture radius

| Anode aperture radius(mm) | Current(A) EGUN | Current(A) CST-PS | Perveance ($\mu$ P) EGUN | Perveance ($\mu$ P) CST-PS | Electric field Maximum (V/m)) |
|---|---|---|---|---|---|
| 35 | 442.22 | 442.221 | 2.13 | 2.131 | $3.16 \times 10^7$ |
| 41 | 447.87 | 447.8769 | 2.16 | 2.162 | $3.55 \times 10^7$ |
| 47 | 430.08 | 430.081 | 2.07 | 2.071 | $3.38 \times 10^7$ |
| 53 | 400 | 400.01 | 1.93 | 1.931 | $3.34 \times 10^7$ |
| 59 | 360 | 360.001 | 1.74 | 1.739 | $3.31 \times 10^7$ |
| 65 | 317.33 | 317.329 | 1.53 | 1.530 | 3.00 x10$^7$ |

Figure 15, shows the comparison of current (A) values at different anode aperture radii, of EGUN and CST-PS, and confirms the results of both codes. Figure 16, shows the beam trajectories at different anode aperture radii of the gun.

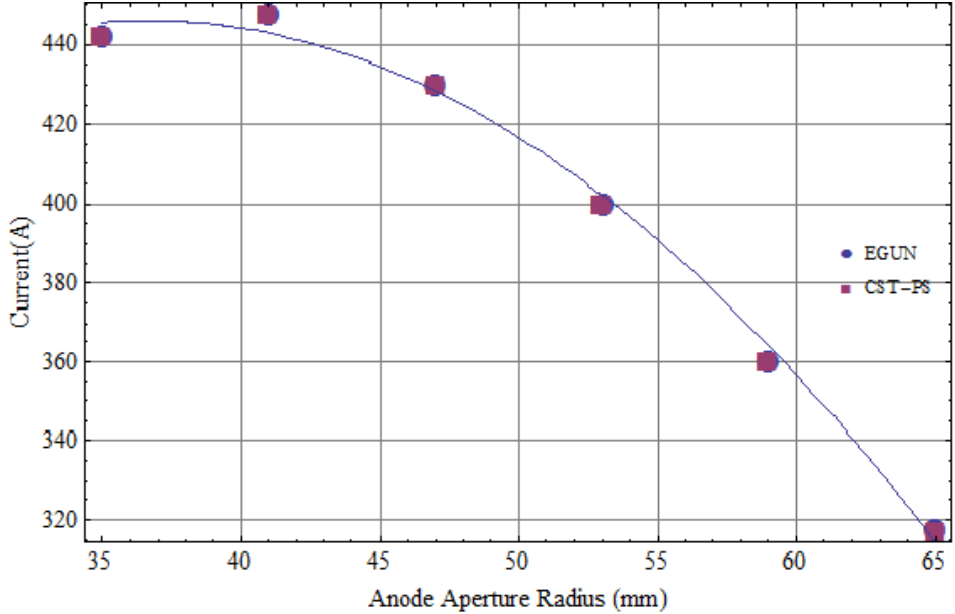

Fig.15: Comparison of current (A) at different anode aperture radii of CST-PS & EGUN

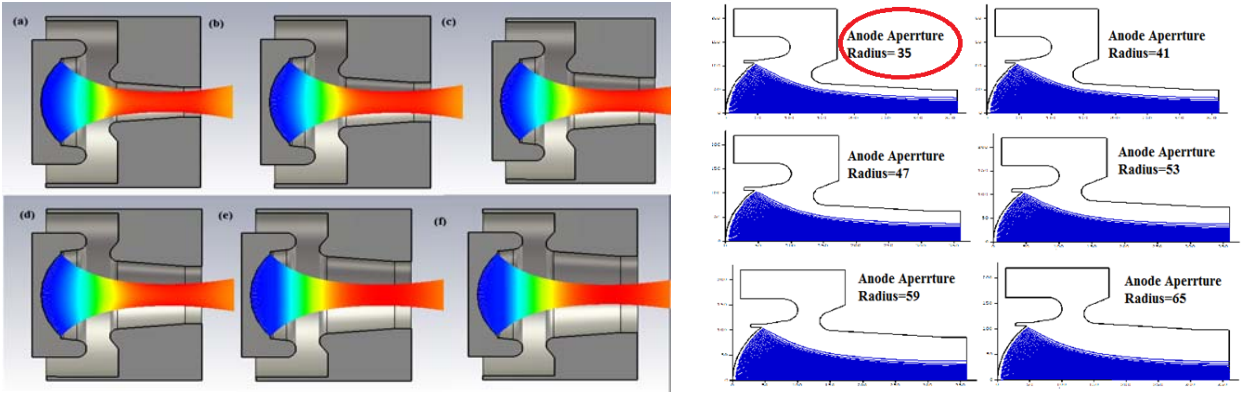

Fig. 16: Beam trajectories at different anode aperture radii: (a) 35 mm , (b) 41 mm, (c) 47 mm, (d) 53 mm, (e) 59 mm, (f) 65 mm

**Anode potential**

The potential difference between the cathode and anode is produced by applying the positive voltage to anode body. In this simulation, the cathode and focusing electrode voltage is kept at the same value. The anode voltage is directly related to beam current. As anode voltage is directly associated to beam current with power of (3/2). However, the perveance is equal to $I/V^{3/2}$ and only depends on the geometry of the electron gun. Therefore, it was expected that current rises with anode voltage but the perveance remains almost constant, the results are summarized in Table 5. Figure 17, shows comparison of beam current of EGUN and CST-PS at differ anode voltages. For this scenario, there will be no change in the beam trajectories.

Table 5: Effects of anode voltage on beam parameters

| Anode Potential (kV) | Current(A) EGUN | Current(A) CST-PS | Perveance (µ P) EGUN | Perveance (µ P) CST-PS | Electric field Maximum (V/m)) |
|---|---|---|---|---|---|
| **250** | 252.6 | 252.6 | 2.02 | 2.02 | $2.26 \times 10^7$ |
| **300** | 330.7 | 330.7 | 2.01 | 2.01 | $2.7 \times 10^7$ |
| **350** | 414.5 | 414.5 | 2.00 | 2.00 | $3.16 \times 10^7$ |
| **400** | 501.9 | 501.9 | 1.99 | 1.99 | $3.6 \times 10^7$ |
| **450** | 597.3 | 597.3 | 1.98 | 1.98 | $4.017 \times 10^7$ |

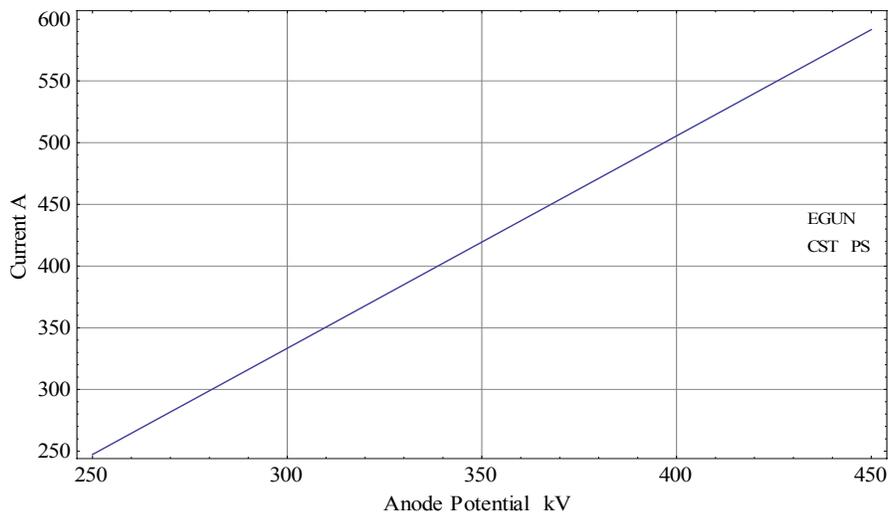

Fig. 17: Comparison of EGUN and CST-PS beam current at different anode voltages

**Conclusions**

Two simulation codes were used to calculate and for verification of the design of 5045 S-band klystron DC electron gun. The gun was optimized to emit 414A beam current and 2.0μP perveance. The emission parameters are in full agreement with the experimental values of the gun. DC gun parameters; beam size, beam emittance, are found compatible with the emitted current and perveance of the beam. Final specifications of the DC gun are reported in Table 6.

Table 6: Results of the DC electron gun

| Acceleration voltage (kV) | 350 |
| Beam current (A) | 414.00 |
| Perveance (μp) | 2.00 |
| Emittance (π mm-mrad) | 103.10 |
| Beam diameter (mm) | 26.82 |
| Beam throw distance (mm) | 157.00 |

The geometrical parameterization provides a deep insight of the structure and dynamics of the beam. As, the 5045 S-Band Klystron RF gun has set standard in linear accelerators technology, therefore the present work is important to disseminate its full range parameters in simplified term as DC beam for vast diverse scientific applications.